# An Analysis of Personal Information Privacy Concerns using Q-Methodology

**Gregg Martin**
Department of Computer Science and Software Engineering
University of Canterbury
Christchurch, New Zealand
Email: gregg.martin@uclive.ac.nz

**Hritik Gupta**
Department of Accounting and Information Systems
University of Canterbury
Christchurch, New Zealand
Email: hritik.gupta@canterbury.ac.nz

**Stephen C. Wingreen**
Department of Accounting and Information Systems
University of Canterbury
Christchurch, New Zealand
Email: stephen.wingreen@canterbury.ac.nz

**Annette M. Mills**
Department of Accounting and Information Systems
University of Canterbury
Christchurch, New Zealand
Email: annette.mills@canterbury.ac.nz

## Abstract

Information privacy has gained increased attention in recent years. This paper focuses on a particular aspect of privacy, i.e., personal information privacy. In this paper a conceptual framework is developed based Westin's theory of Personal Information Privacy (PIP). Concourse theory and Q-methodology was used alongside the literature and the New Zealand Privacy Act 1993 to develop a Q-sort questionnaire. The resulting 29 statements were then sorted by 12 students (majoring in IS Security). The results indicate that for some, privacy priorities may be stable across contexts, and for others this differs, suggesting that current views of privacy (e.g. Westin's theory) may need revising for the modern digital age. The Q-sort methodology also identified three types, each representing distinct collective perspectives on personal information privacy. These types are discussed along with implications and suggestions for future research.

## 1 Introduction

With the explosive growth of online and mobile technologies that thrive on the collection, use and dissemination of personal data, interest in personal privacy is increasing. The situation is exacerbated further by new and emerging technologies such as GPS reporting through cell phones and the Internet of Things which, because of their pervasive and ubiquitous nature, and ability to collect previously uncollectible information, are posing threats to privacy that have hitherto, not been considered (Conger et al. 2013; Ziegeldorf et al. 2014). For the individual, several problems including significant risks to the privacy and security of personal information (Rajindra et al. 2014; Bansal et al. 2015) may arise as the right to control and decide about what personal information is passed on to others is eroded (Smith et al. 2011), or information is stolen, co-opted or compromised (Conger, et al. 2013). In today's context, it is clear that individuals can no longer control their personal information privacy (PIP).

Early research on privacy concerns in the Internet age proposed that PIP was primarily the responsibility of individuals (Smith, et al. 1996). This was based on a long-held view of privacy as the right to be left alone, and that it is the individual's responsibility to maintain that right (Warren and Brandeis, 1890). However, privacy in the modern digital age is now a more complex concept that involves trade-offs between concerns about the collection of personal information and disclosure of such information for some gain (Smith et al. 2011). Privacy is therefore a context-driven concept that is characterised by complex dynamic relationships that must be balanced with the benefits of information sharing (Acquisti et al. 2015; Smith et al. 2011). Recognising this need for balance, Westin (1970) advocated privacy as *the right to define for oneself when, how and to what extent information is released*. Thus, from Westin's viewpoint PIP may be said to relate to identifying the information that an individual wants to keep private. Westin does not address how that information is managed (e.g. who may collect, store and manage that data and for what purpose) nor does he consider the context in which privacy trade-offs are made.

The current state of privacy research fails to capture the full range and richness of issues that are important to people when they make decisions about personal privacy. Westin's metrics, although good and valid for their intended purpose, are relatively one-dimensional, and typically consist of only three or four self-response items (Kumaraguru and Cranor, 2005). Westin's research, over time, led to the development of a spectrum of PIP beliefs, with privacy "*fundamentalists*" at one end of the spectrum, "*unconcerned*" at the other, and "*pragmatists*" in the middle (Kumaraguru and Cranor, 2005). Given the complexity of peoples' privacy beliefs, this raises the question about whether a single, uni-dimensional scale, such as that proposed by Westin, is capable of capturing a complete picture of PIP. To address this question, it is necessary to develop a new means of operationalizing and measuring PIP, that captures the full range of privacy beliefs and how those beliefs are prioritized, and compares the results obtained by those measurements with Westin's theory of PIP.

Westin completed most of his research, and refined his theories in the years before widespread surveillance, collection and data mining of personal information became common. In other words, the global context of PIP has changed dramatically in recent years, and this has changed peoples' beliefs and attitudes with respect to PIP. This alone would necessitate the revisiting of Westin's theory, in light of the new global information context.

Therefore, the research questions addressed by this study are:

RQ1: What are peoples' privacy priorities about how their personal data is used?

RQ2: Do peoples' privacy priorities change with respect to the context (e.g. business vs. social)?

This paper proceeds as follows. First it reviews the literature on the development of PIP research. The paper then describes the research methodology, in which the methods and procedures prescribed by concourse theory are used to develop and pilot test a set of Q-sort items (drawn first from the dimensions identified in prior research and supported in this context by the New Zealand Privacy Act 1993) – the intent is to re-validate, revise, or expand Westin's PIP theory for the modern privacy context. We then present the findings of a pilot study using the new instrumentation. The paper concludes with the relevance of this research and provides some recommendations for further research in the area of PIP.

## 2  Literature Review

The rapid expansion of the Internet through the 1990s, coupled with the growth of social media and mobile technologies, and increasing capacities for data mining and analytics has created new technological contexts that influence privacy concerns differently (Smith et al. 2011). As a result, the concept of personal privacy and what this means must be continuously defined, refined, and re-defined. Early research on the topic was limited mostly to demographic (e.g. name, address, date of birth) and transactional data (Cheung, Chan, and Limayem, 2005; Smith, et al., 2011). Attempts during this time to refine the concept of PIP and move away from one-dimensional views focused on measures of privacy concern. This identified *collection*, *unauthorised use*, *improper access* and *errors* as key dimensions of information privacy concerns (Smith et al. 1996). This was later refined by Malhotra et al. (2004) which proposed the notion of Internet User Information Privacy Concerns (IUIPC) as related to *collection*, *control* and *awareness of privacy practices*. Drawing on this and other works the concept of PIP has evolved to focus on data *collection*, *secondary use*, *ownership/control*, *accuracy*, *awareness* and *access* as key characteristics (Hong and Thong, 2013; Smith, et al. 2011).

A key challenge that arises in the modern digital age is how to protect personal information in a context where a vast spectrum and variety of data, not merely demographic and transactional, are routinely shared. For example, the ability to control what information is revealed on the Internet and who can access, collect, share and otherwise distribute that information is being redefined by changes in technology. The growing use of data mining has also expanded the capacities of firms to collect, store and combine information about individuals. For example, the proliferation of public records on the Internet in the name of 'open government' means that individuals, companies and other agencies can now easily access and combine vast amounts of personal information from various sources on individuals. Similarly, the concept of 'affiliate sharing' of information has allowed merged organisations, or those providing certain services to their business partners to bring together client data in an unprecedented manner. These developments have undoubtedly created new contexts that challenge the prevailing perspectives on notions of "individual privacy" and giving rise to a previously unexplored domain of PIP issues.

For example, drawing on Westin's research, researchers on Internet use have identified a number of issues, concerns, and considerations relating to PIP. Research conducted in the early to mid-2000's focused on where, and how information is collected, and whether the consumer is aware of the collection, including access to personal information, perceived risks of information collection, vendor trust/reputation, theft of personal information, demand for individual control, corporate privacy-related policies and practices, and government surveillance and monitoring of Internet activity (McKnight, et al. 2004; Olivero and Lunt, 2004; Paine et al. 2007; Dinev and Hart, 2006; Dinev et al. 2008). There is also some research that is focused on the taxonomy of PIP concerns, for instance, Kumaraguru and Cranor (2005) examined participant's responses to Westin's various privacy surveys with the intent reviewing and refining the "fundamentalist – pragmatist – unconcerned" spectrum proposed by Westin.

More recent privacy research in areas of e-commerce, healthcare, government interests, and Internet of Things, have uncovered many benefits of information sharing but also show that this new era is producing far more data than before and using different methods to collect and share data, so raising new issues and concerns about PIP (Acquisti et al. 2015; Wright, et al. 2008; Ziegeldorf et al. 2014). Some research has also begun to differentiate between concerns about the flow of personal information between consumer and online agencies, and concerns about how such information is managed by online agencies (Hong and Thong, 2013) and to evaluate the effects of inter-organisational data sharing (Wakefield and Whitten, 2006). As a consequence it makes sense to revisit the early notions of PIP, and what these mean in a new information age.

Westin (1970) defines privacy as the right of individuals to determine for themselves when, how and to what extent information about them is released to others. This conceptualisation of privacy aligns with modern perspectives on the privacy of personal information which focus on what information persons wish to keep private rather than on how this information is managed (Conger, et al. 2013). It also recognises that individuals will provide personal information under a number of circumstances in order to gain some benefit, such as, for personalisation, financial awards, or social benefit (Smith et al. 2011). At the same time these disclosures present challenges when such information is acquired, used, disseminated to others or otherwise compromised without the individual's permission (Conger et al. 2013).

Overall, the literature review revealed that, although there has been extensive research into the types of personal information collected and the elements of information privacy, there is very little to suggest that Westin's theory of PIP has been re-validated for the modern context of PIP. Therefore, this research will proceed to develop a new means to operationalize, measure, and interpret how people conceptualize PIP in the modern context of global information systems, surveillance, social media, "big data", and data mining.

## 3     Research Model

A conceptual framework of determinants of PIP priorities is proposed for this project (Figure 1). The framework is an adaption of Malhotra et al.'s (2004) IUIPC model. It encompasses core elements of individuals' concerns for information privacy (Malhotra et al. 2004; Smith et al. 2011), and key issues that characterise PIP, in this case focusing on disclosure and use (Dhillon and Torkzadeh, 2006; New Zealand Privacy Act, 1993; Westin, 1970). This model is relevant for the goals of this study, since it provides an initial context for a modern concept of PIP, the various issues that are incorporated into individual PIP, and hints at what kinds of measurements ought to be included in a reconceptualised measurement of PIP. The categories represented in Figure 1 will be used as a foundation for developing new instrumentation to capture how individuals prioritize between disclosure, awareness, storage, use, and collection issues, as these pertain to their PIP beliefs.

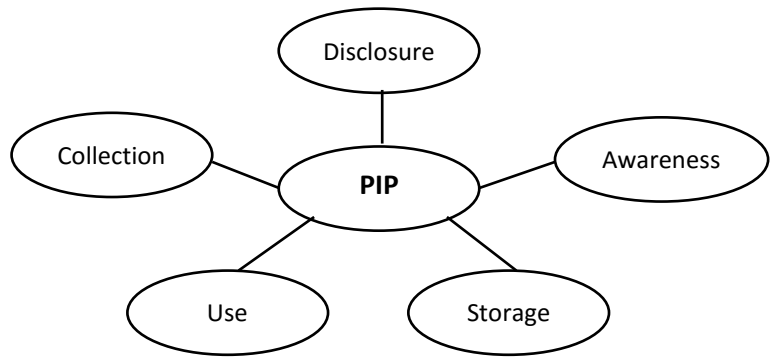

*Figure 1: A Conceptual Model of Personal Information Privacy (adapted from Malhotra et al. (2004) IUIPC Model)*

## 4 Research Method

Concourse theory proposes that people form their belief and value systems within a universe of ideas, feelings, thoughts, and related referential material (Brown, 1980; Stephenson, 1986a; Stephenson, 1986b; Sarkar, et al., 2013; Wingreen, et al. 2009; Wingreen, et al. 2005). A "concourse" is the universe of ideas or statements on any given topic, and a person's belief or value system with respect to the concourse is manifested by how that person prioritizes the ideas and thoughts within the "universe" of the concourse. Q-methodology is the proposed means of operationalising and analysing a concourse, and the person's unique system of beliefs and values with reference to the concourse. To accomplish this, Q-methodology implements the Q-sort to measure, and Q-factor analysis to analyze and interpret the person's belief and value system (Stephenson, 1986 – 1988; Sarkar, et al., 2013; Wingreen, et al. 2005). When performing a Q-sort, the person classifies a field of statements, which are a representative cross-section of the concourse, into a quasi-normal distribution. The distribution is usually formed along a continuum of subjective belief, such as "most important" to "least important", or "most desirable" to "most undesirable". In this manner, the Q-sort operationalises the person's subjective belief or value system with reference to the concourse. Therefore, concourse theory, Q-methodology, and the Q-sort are also appropriate to analyse the structure of people's beliefs systems with reference to the collection, use, and storage if PIP in different online environments. In this manner, RQ1 will be addressed, as the use of Q-methodology to operationalize PIP will make it possible to examine the theorized uni-dimensionality of Westin's PIP scale.

The primary benefit of using q-methodology is that it provides a rich and interpretive understanding of the phenomenon of interest, and it also has advantages in terms of minimal demands on the sample size (Brown, 1980). Application of q-methodology began with the development of q-statements which represented the concourse, in this case, that of the PIP beliefs of individuals in online environments. For this study, the dimensions of PIP representing the concourse were obtained from the literature and corresponding q-statements were drawn from the representation of these dimensions in the New Zealand Privacy Act 1993.

### 4.1 Instrument Development

A Q-sort instrument, comprised of a set of 29 Q-sort statements, was developed according to the guidelines proposed by previous research (Brown, 1980; Stephenson, 1986a; Stephenson, 1986b). Since a concourse may potentially contain an infinite amount of referent material, it is important that a set of q-sort statements be a representative cross-section of the concourse. Toward this goal, there are certain guidelines which, if followed, will yield a high probability that the statements that are selected will be representative of the domain of the concourse. In the context of this study, these guidelines include a review of the relevant literature and the New Zealand Privacy Act 1993, feedback from a focus group comprised of potential study participants, and input from an expert panel to advise on the scope and quality of the sample of Q-statements (Brown, 1980; Sarkar, et al., 2013; Wingreen, et al. 2009). Statement development was also guided by the categories represented in the research model, as presented in Figure 1.

Researchers may also choose to develop a "structured q-set" that represents categories of theoretical interest, by balancing the number of Q-statements in each theoretical category, in a similar fashion to how experimental subjects are assigned to treatment groups for controlled experimental designs in classical correlational research (Watts and Stenner, 2012). To accomplish this, we selected statements the New Zealand Privacy Act 1993 which represent the various dimensions of PIP, which, in theory, should be reflected in the decisions and personal belief systems regarding PIP. Following the instrument development procedure, the final set of 29 statements was judged to be both representative of the larger concourse and well-balanced with regards to the theoretical categories of interest in the current research.

Due to the different environments in which PIP is applicable, two Q-sort scenarios were developed, using the same Q-set. The first scenario asks participants to complete a sorting exercise of each of the 29 statements in relation to their PIP in a social media context. The second scenario asks participants to complete a second Q-sort exercise in relation to their PIP in an online business context. In this way, RQ2 could be addressed, since it supports a test of the null hypothesis that peoples' beliefs are invariant between contexts.

### 4.2 Study Context

Twelve participants were recruited for this research, all of whom were Information Systems students in a New Zealand University, who were also actively involved in an upper-level information security course, which included a great deal of content on information privacy, and other key issues of relevance to PIP. The participants were instructed to look at the 29 q-statements and sort them between 'most important' and 'least important', according to how they felt about their personal information privacy. Each of the 12 participants completed two q-sorts - the first in the context of personal information privacy in social media and, the second sort in the context of personal information privacy and online businesses.

Participants were instructed to sort the statements 'from the outside in', so choosing two statements that were most important to them and two that were least important to them and putting them in the most extreme categories. The participants were instructed to then select four and five statements each for the next two categories of 'most important' and 'least important' statements. This was done to achieve a quasi-normal distribution of the q-statements between the 'most important' and the 'least important'. All unranked statements were categorised as 'neutral'. The collected q-sorts were analysed with the PQ-method software that is commonly used in q-methodology research (Brown1996). The aim of the data analysis was to pilot test the instrument for use in primary data collection in future studies.

As mentioned previously, each participant completed two q-sorts, one for their PIP in the context of online social media, and one for their PIP in the context of online business, so as to investigate RQ2. All q-sorts were labelled according to the PIP context, and all were factored together, so as to make direct comparisons possible when the data are interpreted in the next phase of the research.

## 5 Research Findings

Q-sort data is analysed using a factor analysis, which reveals the factor structure of peoples' perspectives about the concourse. Contrary to the goals of factor analysis in classical statistics, a q-factor analysis factors people into "types", rather than variables into factors. In other words, the result of a q-factor analysis is a typology of peoples' beliefs and priorities about the concourse of interest. This will allow RQ1 to be addressed, since it supports the investigation of whether peoples' perspectives are of one type, as Westin proposes, with "fundamentalists" at one end of the spectrum and "unconcerned at the other, or whether PIP is a more complex and nuanced phenomenon, with multiple perspectives of different types.

Therefore, a centroid factor analysis of two, three and four factor solutions were obtained. The four and three factor solutions were examined as a starting point in the analysis; however, the fourth factor was rejected on account of insignificance. The two-factor solution was inferior to the three-factor solution because it explained less variance in the population of perspectives held by the participants. The three factor solution explained more variance, and minimized the inter-correlation between types, and therefore was accepted. The results of the factor analysis are reported in Table 1 below. Table 1 presents the top 5 and bottom 5 priorities only for each factor – these are emphasised in boldface. The three factors explained 54% of the variance between peoples' perspectives, as represented by their q-sorts. These three types seem to represent the "fundamentalist", and two types of "pragmatist", respectively; the "unconcerned" perspective was not represented in the study.

| Statement | Type 1 | Rank | Type 2 | Rank | Type 3 | Rank |
|---|---|---|---|---|---|---|
| My personal information must not be collected unless it is for a lawful purpose connected with a business function or activity | **1.54** | **4** | -0.12 | 14 | 0.05 | 15 |
| My personal information must not be collected unless it is necessary for the purpose for which it is collected | **1.54** | **4** | -0.43 | 19 | 0.09 | 13 |
| When my personal information is collected, it must come directly from me | -0.62 | 21 | **1.16** | **5** | **2.22** | **1** |
| When my personal information is collected, I must authorise its collection | **1.57** | **2** | **1.84** | **1** | **1.11** | **5** |
| When my personal information is collected, I must be informed of its collection, and the purpose for the collection | **1.57** | **2** | **1.21** | **4** | 0.23 | 11 |
| When my personal information is collected, I must have the rights to access it and be able to correct it | -0.33 | 18 | 0.79 | 7 | **1.5** | **3** |
| My personal information must not be collected using unlawful means, or which intrude unreasonably on my personal affairs | **1.24** | **5** | **1.55** | **3** | **1.28** | **4** |
| An organisation holding my personal information must protect it against loss, unauthorised access, use, modification or disclosure | 0 | 17 | **1.83** | **2** | -0.53 | 20 |
| An organisation that holds my personal information can disclose it to a person, body or organisation if it is necessary to prevent or lessen a serious threat to a person's life | 0.33 | 11 | **-1.61** | **29** | **1.74** | **2** |
| Publicly available personal information about me, does not require my authorisation for collection | **-0.95** | **26** | **-1.30** | **27** | **-1.76** | **29** |
| When my personal information is collected it must not be used in a form that identifies me | **-0.95** | **26** | -0.16 | 16 | 0.19 | 12 |
| If I had given my consent on a recent previous occasion, personal information about me of the same kind can be collected without me being informed of the collection details | 0.62 | 10 | **-1.56** | **28** | -0.4 | 19 |
| I do not need to be informed about the collection of my personal information if my knowledge of the collection does not affect my interests | 0.29 | 13 | **-1.12** | **26** | -0.65 | 21 |
| When an organisation holds my personal information in a way that it can readily be retrieved, I am entitled to have access to it | **-1.86** | **29** | -0.31 | 17 | 0.58 | 7 |
| I am entitled to request corrections to my information held by an organisation, to ensure it is accurate, up-to-date, complete and not misleading | **-1.24** | **27** | -0.15 | 15 | -0.1 | 17 |
| When an organisation holds my personal information, if my request for correction is not made, a "statement of correction sought but not made" must be attached | 0.29 | 13 | -1.01 | 24 | **-1.45** | **27** |
| An organisation that holds my personal information must ensure it is accurate, up-to-date, complete, relevant, and not misleading | **-1.86** | **29** | 0.05 | 12 | -0.21 | 18 |
| An organisation that holds personal information about me that was obtained for one purpose can use it for another purpose if it is publicly available | 0 | 17 | **-1.11** | **25** | -0.88 | 25 |
| An organisation that holds my personal information can use it for purposes other than which it was obtained provided this is directly related to the original purpose | **-0.95** | **26** | -0.42 | 18 | 0.51 | 8 |
| An organisation that holds my personal information can disclose it to a person, body or organisation if this is publicly available information | -0.62 | 21 | -0.72 | 22 | **-1.22** | **26** |
| An organisation that holds my personal information can disclose it to a person, body or organisation if the purpose is directly related to the purpose for which the information was obtained | -0.59 | 19 | -0.70 | 21 | **-1.73** | **28** |

*Table 1. Statement rankings from centroid factor analysis, three-factor solution*

Since all participants provided q-sorts for both online social media, and online business contexts, the distribution of q-sorts between the two contexts was examined, in order to pursue an answer to RQ2. If peoples' perspectives are invariant, as suggested by Westin's research, we would expect each individual's q-sorts to both load on the same factor. Although this was mostly the case, one person, of the twelve, provided q-sorts that loaded on two different factors, suggesting that, at least for some people, their priorities depend on the context. In a larger study size, with more people, or with more precise methodological controls, it should be expected that there will be more types, and more people whose perspectives vary between contexts.

## 6  Discussion and Interpretation

With regard to RQ1, if there were only one spectrum of beliefs, with "fundamentalist" at one end of the scale, and "unconcerned" at the other, it should be manifested as a single factor in the q-factor analysis, with fundamentalists at one pole of the factor, unconcerned at the other, and pragmatists in the middle. Since there were three factors, we find evidence to reject the "uni-dimensional" hypothesis. However, subsequent interpretation revealed that there is nonetheless some merit in the "fundamentalist" and "pragmatist" classifications, even if they are not different perspectives of the same spectrum of belief, since the content of the types 1, 2, and 3 seem to represent the "fundamentalist", and two "pragmatist" perspectives, respectively (further discussion below). Although none of the three types seem to represent an "unconcerned" classification, this may be on account of there being only twelve participants in this pilot study, none of whom represent the "unconcerned" type. To interpret the data in Table 1, a guideline was adopted, that the top 5 and bottom 5 priorities should be examined, since both the highest and lowest priorities are of greatest significance in the definition of a person's perspective, and the results of a q-factor analysis may place a person's perspective at the negative pole of a factor, as well as the positive.

Type 1 (fundamentalist) appears to be an authentication and privacy protection factor and all individuals in this group prefer to give their consent before any action that affects their PIP is initiated. This can be drawn from the most highly ranked statements for Type 1. Type 1 was characterised by the statements "When my personal information is collected, I must authorise its collection", "When my personal information is collected, I must be informed of its collection, and the purpose for the collection", "My personal information must not be collected unless it is for a lawful purpose connected with a business function or activity" and "My personal information must not be collected unless it is necessary for the purpose for which it is collected", "My personal information must not be collected using unlawful means, or which intrude unreasonably on my personal affairs". These priorities point to having a certain limit on the use of personal information by organisations beyond a level that affects the personal information privacy of these individuals. Type 1 individuals can be understood to be 'privacy fundamentalists' with a viewpoint that protecting privacy is the responsibility of the individual himself. Type 1 individuals expect to know the purpose of holding their personal information. This type was least concerned with having access to the collected information.

Type 2 (pragmatist) can be viewed as a protection, and consent to collection of personal information factor. Individuals in this group prefer to be notified of the use of the information before it is collected and expect to be asked for their consent. Such individuals are concerned about the collection of their personal information and think that it should be collected by lawful means and the individual must be notified. The use of their personal information must also be put forward and adequate means must be put in place to protect their information. Accordingly, factor 2 was characterised by the statements "When my personal information is collected, I must authorise its collection", "An organisation holding my personal information must protect it against loss, authorised access, use, modification or disclosure", "My personal information must not be collected using unlawful means, or which intrude unreasonably on my personal affairs." and "When my personal information is collected, I must be informed of its collection, and the purpose for the collection", and "When my personal information is collected, it must come directly from me". Type 2 individuals can be thought of as being a "pragmatist" with a modern view of personal information privacy. In saying that, such individuals believe that information collection is a part of modern web activities like social media and online businesses but a good practice is to associate personal information collection with a clear motive and use behind collecting it. This group was least concerned about collection if consent had been given previously, or if the collection did not impact them.

Type 3 (pragmatist) can be interpreted as a consent and transparency factor. Such individuals believe that any collection of personal information must be accompanied by prior consent and any collected information can be disclosed to another person or organisation if this is deemed essential. However, transparency underlies the collection of any personal information, and such information must be available to the type 3 person. Factor 3 was characterised by the statements "When my personal information is collected, it must come directly from me", "An organisation that holds my personal information can disclose it to a person, body or organisation if it is necessary to prevent or lessen a serious threat to a person's life", "When my personal information is collected, I must have the rights to access it and be able to correct it", "My personal information must not be collected using unlawful means, or which intrude unreasonably on my personal affairs", and "When my personal information is collected, I must authorise its collection". Such individuals possess characteristics of Type 2 individuals which are reflected by a sense of modern pragmatism in regards to collection of personal information and are also open to the use of their information for meaningful purposes. This group was least concerned with the sharing of publicly available information, or if the sharing of collected information is for a related purpose.

With regard to RQ2, the results of the factor analysis also reveal that for most of the individuals, except one, both of the social media and online business q-sorts loaded on the same factor, which indicates that the priorities and beliefs of the individuals are similar in both the cases. This finding mostly supports Westin's theory, although it should be noted that it also provides evidence that Westin's theory is not complete, since there are some people whose perspectives do vary from one context to the next. There is also a question as to whether the participants whose perspectives were invariant in this study, would remain invariant if other contexts are examined, as only online social media and online business contexts were examined in this study. It is also possible that the methods employed in this study, specifically that both q-sorts were obtained in immediate succession, lacked the precision and granularity to distinguish between-context variance for some individuals. Also, since even in a small sample of n = 12, we found evidence of invariance, we would expect there to be even more in a larger study with more participants. In summary, our results support the need to expand Westin's theory to larger, more complex, and more modern contexts of PIP.

## 7  Conclusion

We have identified a gap in research related to personal information privacy in online environments. Accordingly, we conducted an extensive literature review to identify what were the main aspects of PIP that people identified with most and undertook an empirical study using Q-sort instrumentation to analyse these aspects and whether there were differences across contexts. This study contributes to the limited body of knowledge on PIP priorities in three ways. Firstly, we have developed and pilot tested an instrument to capture privacy priorities. Secondly, this is the first attempt to 'type' the belief systems of individuals with regards to their PIP and to understand their privacy concerns related to the collection, storage, and use of personal information in an online environment. Thirdly, this research finds sufficient evidence to pursue a revision and expansion of Westin's theory of PIP in the context of modern information systems.

Although the current study is small in size and scope, it provides a starting point for further research into belief systems of PIP. It should be noted that the data collection was situated within New Zealand, using student respondents who had been exposed to a great deal of content on key concepts and issues related to PIP; as such their views may not reflect the opinions and priorities of a more general population. For instance, we found no evidence in this study for an "unconcerned" type, although we may expect to find "unconcerned" people in a larger study of a more general population. Second, while the selected dimensions of PIP were informed by the literature, the q-statements were based on corresponding statements in the New Zealand Privacy Act 1993 which allowed for the situating and contextualising of PIP beliefs. Although some might argue that the use of statements drawn from the New Zealand Privacy Act could be a limitation, it should be noted that the statements also correspond to the dimensions of PIP in the literature. New Zealand is widely regarded to be a global leader in the field of personal freedom and privacy, and therefore the law, as it is written, is fairly representative of the public concourse in a nation that is well-informed about personal privacy issues. In other words, the public discourse about information privacy in New Zealand may an ideal environment from which to sample a concourse on the topic of PIP, and a counter-argument could be made that the use of NZ law is actually a strength of the study, rather than a limitation. Finally, although this study found evidence that at least some peoples' PIP beliefs vary between contexts, most people were invariant. This suggests the possibility that methodological precision might be an important issue. Future research should seek methodological improvements that will yield more precise results with regard to peoples' between-context variance in PIP beliefs. A larger, more comprehensive study is recommended, which should reveal a deeper and even more complex picture of PIP, than the one demonstrated here, which included only twelve participants, providing q-sorts for two different contexts.

There are a number of avenues of future research, including examining a greater range of participants. Future empirical research should also attempt to understand relationship of PIP concerns with other aspects of data collection, such as trust and risk (Malhotra et al. 2004). Finally, the results have implications both for researchers who are looking for theories that explain the importance of understanding PIP concerns with future technology and commercial companies who may look to address some of the concerns identified in this research as these relate to the way in which they collect, manage and use peoples' personal information.